\documentclass[aps,pre,amsfonts,amsmath,twocolumn,groupaddress]{revtex4-1}
\pdfoutput=1 
\usepackage{graphicx}
\usepackage{amssymb}
\usepackage{amsmath}
\usepackage{amsfonts}
\usepackage[usenames]{color}
\usepackage{gensymb}
\usepackage{multirow}
\usepackage{amsbsy}
\usepackage{epsfig}

\newcommand{\be}{\begin{equation}}
\newcommand{\ee}{\end{equation}}
\newcommand{\bea}{\begin{eqnarray}}
\newcommand{\eea}{\end{eqnarray}}
\newcommand{\ben}{\begin{enumerate}}
\newcommand{\een}{\end{enumerate}}
\newcommand{\bit}{\begin{itemize}}
\newcommand{\eit}{\end{itemize}}

\numberwithin{equation}{section}

\definecolor{BrickRed}{cmyk}{0,0.89,0.94,0.28}
\definecolor{MidnightBlue}{cmyk}{0.98,0.13,0,0.43}
\definecolor{DarkGreen}{rgb}{0,0.7,0.1}
\newcommand{\dif}{\mathrm{d}}

\usepackage{ulem}

\begin{document}

\title{Scattering Approach for Fluctuation--Induced Interactions at Fluid Interfaces}
 
 \author{Ehsan Noruzifar}
 \author{Jef Wagner}
\author{Roya Zandi}
 \affiliation{Department of Physics and Astronomy,
   University of California, Riverside, California 92521, USA}

\begin{abstract} 
We develop the scattering formalism to calculate the interaction between colloidal particles 
trapped at a fluid interface. Since, in addition to the interface, the colloids may also fluctuate in this system, we implement the 
fluctuation of the boundaries into the scattering formalism and investigate how
the interaction between colloids is modified by their fluctuations. 
This general method can be applied to any number of colloids with various geometries at 
an interface. We apply the formalism derived in this work to a system of spherical colloids at 
the interface between two fluid phases. For two spherical colloids, this method 
very effectively reproduces the previous known results. For three 
particles we find analytical expressions for the large separation asymptotic energies and numerically calculate the Casimir interaction at all separations.
Our results show an interesting three body effect for fixed and fluctuating colloids. While the three body effect strengthens the attractive interaction between fluctuating colloids, it diminishes the attractive force between colloids fixed at an interface.
\end{abstract}

\maketitle

\section{Introduction}
\label{introduction}

Colloidal particles at a fluid interface can form complex two
dimensional (2D) patterns \cite{oettel_la,oettel_condmat}.  A deep
understanding of different interactions between the colloids trapped
at the interface is necessary to explain the observed patterns.  Among
the forces present at the interface, the importance of
fluctuation-induced forces has become more apparent recently, due to
the advancement of technology for the detection of these forces at
nano scale \cite{beschinger_epl,beschinger_pre,beschinger_nature}.

Fluctuation-induced forces were first predicted by Casimir in 1948
\cite{casimir} who showed that in vacuum, confinement of quantum
fluctuations of the electromagnetic fields by two parallel uncharged
conductors gives rise to an attractive force between them.  Since
then, several experiments have measured these forces with high
precision \cite{lamor,mohideen,umar2000,decca1,decca2}. Quite interestingly, this effect is not
restricted to electromagnetic systems and can be generalized to any
fluctuating field in which the fluctuations are modified by the
presence of external objects.  For example, Casimir-like forces exist
between particles embedded in a membrane or trapped at a fluid
interface, which restrict thermal height fluctuations of the membrane
or fluid interface. Indeed, the fluctuation-induced forces in soft matter systems have been subject of intense research for several
decades now \cite{mehran_modern,kardar_gol,bruinsma,mat}.

About 20 years ago, fluctuation-induced forces were investigated
between thin rods on a film \cite{kardar_gol}
through a path integral formalism for manifolds \cite{kardarli}.
Later the formalism was used to study the fluctuation-induced forces
between disks and spherical colloids \cite{oettel06,oettel07} as well
as ellipsoidal colloids trapped at a fluid interface between two fluid
phases \cite{mine_epje, mine_pre}.  In a recent work, an effective
field theory method has been employed to find the asymptotic expansion
of the fluctuation-induced interaction \cite{eft_deserno1,
  eft_deserno2}. Around the same time, Lin {\it et al.} presented a
Green's function method to calculate the fluctuation-induced forces
between soft and hard particles embedded in a membrane \cite{roya}.

In this work, we study the interaction between colloidal particles due
to the confinement of thermal height fluctuations of the interface.
Since the shape and degrees of freedom of the fluctuations contribute
significantly to the interactions between the colloids at the
interface, we use the scattering formalism
\cite{universal,emigScalar}, which has proven to be very effective for
the calculation of the Casimir forces between many objects with
different geometries and material properties.  The method is also
extremely efficient for numerical calculations, especially at short
separation.

The scattering method has been widely and successfully used for the
calculation of the QED Casimir forces \cite{roya10,
  ours1,rahi08,emig06,ours,Emig:2009fk,Graham:2010uq,Maghrebi:2011kx}.
While in all the QED cases the position of the external objects are
assumed to be fixed in space, for the colloids trapped at an interface
both the object and interface fluctuate. To this end, the most
significant aspect of this work is to include the colloid fluctuations
into the scattering formalism.
 
The scattering method simplifies the fluctuation-induced problems by
separating the calculation into finding the translation matrices
($\mathbb{U}$) and scattering matrices or $T$-matrices
($\mathbb{T}$). The $\mathbb{U}$ matrix corresponds to the way the
fluctuations propagate through the field between the objects and
the $T$-matrix represents the interaction of the object with the
fluctuations. This work further simplifies the problem by separating
the way a particle interacts with fluctuations from the type of
fluctuations the particle undergoes. 

We apply this
modified scattering method to calculate the interaction between
spherical Janus particles.  In particular, we study three different
types of particle fluctuations: colloids frozen at a vertical
position, bobbing colloids that fluctuate only vertically with the
interface, and bobbing and tilting colloids that both fluctuate
vertically and tilt side to side.  We are able to easily reproduce the
previously known results between two spherical Janus particles. We
also calculate the fluctuation-induced interaction between three
particles and find that both bobbing only and bobbing and tilting
particle fluctuations weaken the attractive interaction between
particles. In contrast, the three body effect makes the attractive
Casimir force stronger between frozen particles.

The structure of the paper is as follows: We present the general form
of the Hamiltonian for colloids trapped at a fluid interface in
Sec.~\ref{sec:hamilton}, and introduce the partition function of the
colloids trapped at the interface in
Sec.~\ref{sec:partition-function}. The corresponding partition function in terms of
the scattering parameters is evaluated in Sec.~\ref{sec:pf-eval}. Using the derived
partition function, the general form of the Casimir energy for
colloids with arbitrary geometry is obtained in
Sec.~\ref{sec:casimir-energy}, and the application of this method to a
system of spherical janus colloids is presented in
Section~\ref{sec:numerics}. A summary of this work and our main
conclusions are presented in Sec.~\ref{sec:summary}. Note that a brief
introduction to this method was published by the authors in
Ref.\cite{fluid_letter}.


\section{Hamiltonian}
\label{sec:hamilton}
We consider an infinite interface between two fluid phases,
characterized by the surface tension $\sigma$.  Due to the thermal
fluctuations in the system, the interface deviates from its
equilibrium flat position placed at $z=0$.  We assume that the thermal
height fluctuations are small, without overhangs and bubbles.  To this
end, we parameterize the interface height profile with the Monge
representation, i.e. $z=u({\bf x})$.  The free energy costs associated
with the interface fluctuation is then
\begin{equation}
\label{eq:hamiltonian_interface}
\mathcal{H}_{\text{int}} [u] = \sigma\int_{\mathbb{R}^2} \dif^2 x 
\Big[ \sqrt{1+\big(\nabla u)^2} - 1 + 
  \frac{\Delta \rho g}{2\sigma} u^2 \Big]\,,
\end{equation}
where the first two terms represent the energy costs associated with
the height fluctuations and the third term represents the change in
the gravitational energy.  Since the thermal fluctuation does
not create a huge change in the interface profile, we perform a small
gradient expansion ($|\nabla u|\ll 1$) in
Eq.~\eqref{eq:hamiltonian_interface} and obtain the capillary wave
Hamiltonian
\begin{equation}
\label{eq:capillary-wave-hamiltonian}
\mathcal{H}_{\text{int}}[u] \approx \mathcal{H}_{\text{cw}}[u] =
\frac{\sigma}{2}\int_{\mathbb{R}^2} 
\dif^2 x\,\left[  \big( \nabla u\big)^2 
+ \frac{u^2}{\lambda_{\rm c}^2}\right]\,,
\end{equation}
with $\lambda_{\rm c} = \sqrt{{\sigma}/{(\Delta \rho g})}$ the
capillary length.  Minimizing the interface Hamiltonian with respect
to the height $u$, subject to the boundary condition $u=0$ as $x$ goes
to $\infty$, we find the capillary wave Helmholtz equation,
\begin{equation}
\label{eq:helmholtz}
(-\nabla^2 +\lambda_{\rm c}^2)u({\bf r})=0\,.
\end{equation}
We will discuss the solutions of the above equation in the next
section.  

For each colloid we must then add the following term to the total Hamiltonian
\begin{multline}
\label{eq:colloid_hamiltonian}
\mathcal{H}_{\text{col}}^i [u,f_i] = \\
-\sigma \bigg( \int_{\Omega_i} \dif^2 x
\sqrt{1+\big(\nabla u\big)^2} - 
\int_{\Omega_i^{\text{ref}}} \dif^2 x \bigg)\\
-\frac{\Delta \rho g}{2}\bigg(\int_{\Omega_i}\dif^2 x\,u^2 - 
h_i^2\Omega_i\bigg) \\
+ (\sigma_{i,\text{I}} - \sigma_{i,\text{II}}) 
\Delta A_{i,\text{I}},
\end{multline}
with $\Omega_i$ the
projected area of the interior of the colloid, $h_i$ height
of the center of mass of the colloid with respect to its equilbrium
position, $\sigma_{i,\text{I(II)}}$ the surface tension between the
colloid and the fluid in phase $\text{I(II)}$, and $\Delta
A_{i,\text{I}}$ the change in the contact area between the colloid
and fluid phase $\text{I}$. Note that all these parameters are
functionals of the contact line between the interface and the colloid,
represented by the field $f_i$.  The second line in Eq.~\eqref{eq:colloid_hamiltonian} is from the exclusion of energy associated with
the interior regions of the colloid $i$ from the fluid-fluid inteface,
the third line is the change in the gravitational potential energy,
and the fourth line is the added fluid-colloid surface energy.  
Performing a small gradient expansion ($|\nabla u|\ll 1$) in Eq.~\eqref{eq:colloid_hamiltonian}, we find
\begin{multline}
\label{eq:colloid_hamiltonian-mid}
\mathcal{H}_{\text{col}}^{i} [u,f_i] = 
-\frac{\sigma}{2} \int_{\Omega_i} \dif^2 x
\left[\big(\nabla u\big)^2+\frac{u^2}{\lambda_{\rm c}^2}\right] \\
+\frac{\sigma}{2}\Omega_i
\left[\frac{h_i^2}{\lambda_{\rm c}^2}\right] \\
+ \sigma {\Delta\Omega_i} 
+ (\sigma_{i,\text{I}} - \sigma_{i,\text{II}}) 
\Delta A_{i,\text{I}}\,,
\end{multline}
%
with $\Delta\Omega_i = \Omega_i^{\text{ref}}-\Omega_i$ the change in the
projected area of the colloid
from its equilibrium position.  Note that the second line has a negligible contribution in the colloids
free energy and will be dropped as the fluctuations are small and as such $h_i/\lambda_{\rm c}
\lll 1$. The first term can be integrated by
parts to give a correction Hamiltonian to the interface term
\begin{equation}
  \label{eq:correction_hamiltonian}
  H_{\text{cor}}^i = -\frac{\sigma}{2} \! \int_{\Omega_i} \!\!\!\! \dif^2 x \,
  u\big(-\nabla^2 + \lambda^{-2}_{\text{c}}\big) u,
\end{equation}
and the boundary term from the integration by parts can be combined
wih the remaining terms to give a boundary Hamiltonian
\begin{equation}
  \label{eq:boundary_hamiltonian}
  H^i_{b} = -\frac{\sigma}{2} \! \int_{\delta \Omega_i} \!\!\!\!\!\! \dif x \,
  (\hat{n} \cdot \nabla u) u + \sigma \Delta \Omega_i +
  (\sigma_{i,\text{I}} -\sigma_{i,\text{II}})\Delta A_{i,\text{I}}
\end{equation}
%
%
%
The total Hamiltonian of
the interface and colloids system is then given by,
\begin{multline}
 \label{eq:total-hamiltonian}
 \mathcal{H}_{\text{tot}}[u,f_1,\dots,f_N] = 
 \mathcal{H}_{\rm int}[u] \\+ \sum_{i=1}^N 
 \big(\mathcal{H}_{\text{cor}}^i[u]+
 \mathcal{H}_{b}^i[f_i] \big).
\end{multline}
We use Eq.~\eqref{eq:total-hamiltonian} to find the partition function
and the Casimir energy in the colloidal system at the fluid interface.

\section{Partition Function}
\label{sec:partition-function}
In this section we find the partition function of the canonical
ensemble of colloidal particles trapped at the interface between two
fluid phases.  Using Eq.~\eqref{eq:total-hamiltonian}, the partition
function at temperature $T$ is
\begin{equation}
\label{eq:partition-function}
\mathcal{Z} = \int_{\mathcal C} \mathcal{D} u \prod_{i=1}^N \mathcal{D} f_i \exp \left[-\frac{{\mathcal H}_{\text{tot}}}{k_B T} \right]\,,
\end{equation}
where $\mathcal C$ denotes the constraint imposed by the fact that the
interface height field $u$ matches the contact line fields $f_i$.  By
defining the auxiliary fields $\psi_i({\bf x})$ on the boundary of the
projected areas $\delta \Omega_i$, the constraint can be replaced with
a Dirac delta functional
\begin{equation}
\label{eq:dirac-delta-functional}
\delta[u \!-\! f_i] = \frac{k_B T}{\sigma} \!\! \int\! \mathcal{D} 
\psi_i \exp \left[ \imath \frac{\sigma}{k_B T}
\! \int_{\delta \Omega_i} \!\!\!\!\! \dif x\, \psi (u \!-\!f_i) \right]\!.
\end{equation}
Inserting Eq.~\eqref{eq:dirac-delta-functional} into
Eq.~\eqref{eq:partition-function}, we have
\begin{equation}
 \label{eq:partition_int_col}
 \mathcal{Z}= \int \prod_{i=1}^N{\mathcal D}\psi_i \,{\mathcal Z}_{\rm int}\,
 {\mathcal Z}_{\rm col}\,,
\end{equation}
where
\begin{multline}
 \label{eq:partition-int}
 {\mathcal Z}_{\rm int} = \int {\mathcal D}u 
 \exp\left[-{\frac{\mathcal{H}_{\text{int}}[u]+
       \sum_{i=1}^N \mathcal{H}_{\text{cor}}^i[u]}{k_B T}}\right.\\
   \left. +\imath\frac{\sigma}{k_B T}\sum_{i=1}^{N}
   \int_{\delta \Omega_i} \!\!\!\! \dif x\,
   u\,\psi_i\right]\,,
\end{multline}
and
\begin{multline}
\label{eq:partition-col}
{\mathcal Z}_{\rm col}= \int \prod_{i=1}^N{\mathcal D}f_i 
\exp\left[-\sum_{i=1}^N{\frac{{\mathcal H}_b^i[f_i]}{k_B T}}\right.\\
\left.-\imath\frac{\sigma}{k_B T}\sum_{i=1}^N
\int_{\delta \Omega_i} \!\!\!\! \dif x\, 
f_i\,\psi_i \right]\,.
\end{multline}
Equations \eqref{eq:partition_int_col}, \eqref{eq:partition-int}, and \eqref{eq:partition-col}  reveal that we can separate the partition function into the interface and colloid parts, ${\mathcal
  Z}_{\text{int}}$ and ${\mathcal Z}_{\text{col}}$.

The integration over the field $u$ in the interface partition function
$\mathcal{Z}_{\text{int}}$ can be done easily, and we find

\begin{equation}
 \label{eq:partition-int-gij}
 \mathcal{Z}_{\text{int}} = C_0\, 
 \exp\left[-\frac{\sigma}{2k_B T}\sum_{i,j=1}^N G_{ij}\right]\,,
\end{equation}
with $C_0$ a constant and 
\begin{equation}
\label{eq:Gij} 
G_{ij} = \int_{\delta \Omega_i} \!\!\!\! \dif{x} 
\int_{\delta \Omega_j} \!\!\!\! \dif{x}' 
\psi_i({\bf x})\, G({\bf x}, {\bf x}')\,\psi_j({\bf x}')\,,
\end{equation}
where $G({\bf x},{\bf x}')$ is the Green's function of the capillary
wave Hamiltonian for the free interface. The Green's function is
unique up to an homogeneous solution, which we can specify by
demanding that the Green's function be regular on the interior of the
colloids $\Omega_i$.

\section{Evaluating the Partition Function}
\label{sec:pf-eval}
In this section, we use the scattering approach 
to evaluate the partition function, \ref{eq:partition-int-gij}.  The
scattering approach relies on expanding the fluctuating fields in a
complete set of solutions to the corresponding wave equation in order
to evaluate the partition function. This section proceeds in the
following manner: the integrals of the Green's function in
Eq.~\eqref{eq:Gij} are evaluated for $i\ne j$ when the integration
takes place on different colloids, and for $i = j$ when the integration is
over a single colloids. The boundary matrix $H^i_b$ is then
defined by expanding the boundary Hamiltonian to lowest order.

Before proceeding it is necessary to define some properties of the
solutions of the capillary wave equation.  For each colloid, space can
be divided into interior $x\in \Omega_i$ and exterior $x \notin
\Omega_i$ regions. The capillary wave equation will have two sets of
independent solutions: $\phi^{\text{inc}}_{i\alpha}(x)$ regular over
the interior and $\phi^{\text{sct}}_{i\alpha}(x)$ regular over the
exterior. The index $\alpha$ is a separation constant if the solutions
are found through separation of variables, and is the same for both
incident and scattered solutions.

Both sets of solutions are complete and orthonormal over the space
$\delta \Omega_i$, so any function defined on $\delta \Omega_i$ can be
written as an expansion in terms of the incident solutions. To this
end, we will expand both $\psi_i(x)$ and $f_i(x)$ as
\begin{align}
  \psi_i(x) &= \sum_{\alpha} \Psi_{i\alpha} 
  \phi^{\text{inc}}_{i \alpha}(x), \label{eq:psi_expansion}\\
  f_i(x) &= \sum_{\alpha} P_{i\alpha} 
  \phi^{\text{inc}}_{i \alpha}(x), \label{eq:f_expansion}
\end{align}
where $\Psi_{i\alpha}$ and $P_{i\alpha}$ are the expansion
coefficients. We employ the expansions of $\psi_i(x)$ and $f_i(x)$ in
Eqs.~\ref{eq:psi_expansion} and \ref{eq:f_expansion} to find the
Green's function, Eq.~\eqref{eq:Gij}, used in the calculation of the
interface partition function $\mathcal{Z}_{\text{int}}$,
Eq.~\ref{eq:partition-int-gij}.

\subsection{Interface Part}
The $G_{ij}$ term given in Eq.~\ref{eq:partition-int-gij} for the
interface partition function $\mathcal{Z}_{\text{int}}$ can be
simplified using Eqs.~\eqref{eq:Gij} and\eqref{eq:psi_expansion}
\begin{equation}
  G_{ij} = \sum_{\alpha} \Psi_{i\alpha} M^{ij}_{\alpha \beta} \Psi_{j\beta}.
\end{equation}
where $M^{ij}_{\alpha \beta}$ is given by the integral
\begin{equation}\label{eq:Mij_def}
  M^{ij}_{\alpha \beta} = \int_{\delta \Omega_i} \!\!\!\! \dif x
  \int_{\delta \Omega_j} \!\!\!\! \dif x' \,
  \phi^{\text{inc}}_{i\alpha}(x) G(x,x') \phi^{\text{inc}}_{j \beta}(x').
\end{equation}
Note that the indices $i$ and $j$ refer to the auxiliary field defined
on the objects $i$ and $j$. Two different cases need to be considered:
$i \ne j$ and $i=j$.


%
\subsubsection{Interaction terms: \em i $\bf \ne$ j}
For the interaction terms if $i\ne j$, then $x$ and $x'$ are on different
colloids. In that special case the free Green's function can be
explicitly written as a product of incident and scattered solutions
\begin{equation}\label{eq:greens_function_def_1}
  G(x,x') = \sum_\alpha c_\alpha \phi^{\text{inc}}_{i\alpha}(x)
  \phi^{\text{sct}}_{i\alpha}(x'),
\end{equation}
with $x \in \Omega_i$ and $x' \notin \Omega_i$. The quantity
$c_\alpha$ is the expansion coefficient. The integral in
Eq.~\eqref{eq:Mij_def} is then given by the product of two integrals
\begin{multline}\label{eq:Mij_inej_integral}
  M^{ij}_{\alpha\beta} = \sum_\gamma c_\gamma \int_{\delta \Omega_i}
  \!\!\!\! \dif x \, \phi^{\text{inc}}_{i\alpha}(x)
  \phi^{\text{inc}}_{i \gamma}(x) \\ \times \int_{\delta \Omega_j}
  \!\!\!\! \dif x' \, \phi^{\text{sct}}_{i\gamma}(x')
  \phi^{\text{inc}}_{j \beta}(x'),
\end{multline}
with $c_\gamma$ the expansion coefficient.
%
In order to evaluate the second integral, we use the property that the
incident solutions $\phi^{\text{inc}}_{i\beta}(x)$ are complete on the
contact line of particle $j$ and write the scattered solution from
particle $i$, $\phi^{\rm sct}_{i\alpha} ({\bf x}')$, as
\begin{equation}
\label{eq:translate_expansion}
 \phi^{\rm sct}_{i\alpha} ({\bf x}') =
 \sum_\lambda {\mathbb U}_{\alpha\lambda}^{ij}
 \phi^{\rm inc}_{j\lambda}({\bf x}')\,.
\end{equation}
The elements of the translation matrix, $ {\mathbb
  U}_{\alpha\lambda}^{ij}$, are the expansion coefficients of the
scattered solution from particle $i$ in terms of the incident solution
for particle $j$.  Inserting Eq.~\eqref{eq:translate_expansion} into
Eq.~\eqref{eq:Mij_inej_integral} and using the orthonormality of the
solutions, one immediately finds the $M$-matrix elements in terms of
the elements of translation matrix, $ {\mathbb
  U}_{\alpha\lambda}^{ij}$
\begin{equation}\label{eq:Mij_matrix_elements}
M^{ij}_{\alpha_\beta} =  c_\alpha {\mathbb U}_{\alpha\beta}^{ij}.
\end{equation}
The calculation of the elements of $M$-matrix is more complex in case
of self-interaction, $i=j$.
%

%
%
%
\subsubsection{Self-interaction terms: \em i $\bf =$ j}
We proceed with the evaluation of self-interaction terms by defining
the function,
\label{sec:self-interaction}
\begin{equation}\label{eq:gialpha_integral_def}
  g_{i\alpha}(x') = 
  \int_{\delta \Omega_i} \!\!\!\! \dif x \,
  \phi^{\text{inc}}_{i\alpha}(x) G(x,x'),
\end{equation}
over the two different regions,
\begin{equation}
  g_{i\alpha}(x') = \begin{cases}
    g^{\text{in}}_{i \alpha}(x') &\quad\text{for $x' \in \delta \Omega_i$} \\
    g^{\text{out}}_{i\alpha}(x') &\quad\text{for $x' \notin \Omega_i$}
  \end{cases},
\end{equation}
Note that on the contact line, $x' \in \delta \Omega_i$, and outside
the colloid, $x' \notin \Omega_i$. We can now calculate the matrix
elements, $M^{ii}_{\alpha \beta}$, given in Eq.~\ref{eq:Mij_def} as an
integral over the contact line
\begin{equation} \label{eq_mii}
  M^{ii}_{\alpha \beta} = \int_{\delta \Omega_i} \!\!\!\! \dif x \,
  g^{\text{in}}_{i\alpha}(x) \phi^{\text{inc}}_{i\beta}(x).
\end{equation}
Using  Eq.~\ref{eq_mii} and the orthonormality property of the $\phi^{\text{inc}}_{i\alpha}$
functions, the $g^{\text{in}}_{i\alpha}$ function can be expanded on the contact line as
\begin{equation} \label{eq_gin}
  g^{\text{in}}_{i\alpha}(x) = 
  \sum_\beta M^{ii}_{\alpha \beta} \phi^{\text{inc}}_{i\beta}(x)\,.
\end{equation}
Further, the $g^{\text{out}}_{i\alpha}$ function
defined outside the colloid can be written as 
\begin{equation}\label{eq_gout}
  g^{\text{out}}_{i\alpha}(x) = g^{\text{in}}_{i\alpha} + 
  \Delta g_{i\alpha}(x).
\end{equation}
with $\Delta g_{i\alpha}$ a correction term that can be expanded in terms of the solutions to the
capillary wave Hamiltonian
\begin{equation}
  \label{eq:Deltag_expansion}
  \Delta g_{i\alpha}(x) = \sum_\beta \chi^i_{\alpha \beta} \bigg(
  \phi^{\text{inc}}_{i\beta}(x) + \sum_{\gamma} \mathbb{T}^{i}_{\beta
    \gamma} \phi^{\text{sct}}_{i \gamma}(x) \bigg).
\end{equation}
The expansion is allowed because the correction term $\Delta g_{i\alpha}$
can be defined as the difference of the $g^{\text{out}}_{i\alpha}(x)$
defined on the exterior and the $g^{\text{in}}_{i\alpha}$ function
defined on the boundary. The $g^{\text{out}}_{i\alpha}$ function is a
solution to the homogeneous capillary wave equation and can be expressed
in terms of $\phi^{\text{sct}}_{i\alpha}$ and
$\phi^{\text{inc}}_{i\alpha}$. The $g^{\text{in}}_{i\alpha}$ function
is defined on the boundary $\delta
\Omega_i$, and the incident field $\phi^{\text{inc}}_{i\alpha}$ are
complete on the boundary. Note that the correction term must go to
zero on the boundary, and therefore, we have 
\begin{equation}\label{eq:scat_matrix_def}
  \phi^{\text{inc}}_{i\beta}(x) + \sum_{\gamma}
  \mathbb{T}^{i}_{\beta \gamma} \phi^{\text{sct}}_{i \gamma}(x) = 0,
\end{equation}
which defines regular $T$-matrix as the Dirichlet scattering matrix.

We now insert Eqs.~\eqref{eq:Deltag_expansion} and \eqref{eq_gin} into Eq.~\eqref{eq_gout} and consider the fact that the $g^{\text{out}}_{i\alpha}$ function must remain regular in the
exterior of the particle. This
lets us relate $\chi^{i}_{\alpha \beta}$ term in
Eq.~\eqref{eq:Deltag_expansion} to the $M^{ii}_{\alpha \beta}$ matrix
coefficients
\begin{equation}
  \chi^i_{\alpha \beta} = - M^{ii}_{\alpha \beta}.
\end{equation}
Thus the function $g^{\text{out}}_{i\alpha}(x)$ given in Eq.~\eqref{eq_gout} can be written as
\begin{equation}\label{eq:gout_def_1}
  g^{\text{out}}_{i\alpha}(x) = -\sum_{\beta \gamma}
  M^{ii}_{\alpha \beta} \mathbb{T}^i_{\beta \gamma}
  \phi^{\text{sct}}_{i\gamma}(x).
\end{equation}
A second definition for the $g^{\text{out}}_{i\alpha}$ function can be
obtained inserting Eq.~\eqref{eq:greens_function_def_1} into
Eq.~\eqref{eq:gialpha_integral_def}
\begin{equation} \label{eq_gout2}
  g^{\text{out}}_{i\alpha}(x') = \sum_\beta c_\beta
  \int_{\delta \Omega_i} \!\!\!\! \dif x \,
  \phi^{\text{inc}}_{i\alpha}(x) \phi^{\text{inc}}_{i\beta}(x) \,
  \phi^{\text{sct}}_{i\beta}(x').
\end{equation}
Using the orthonormality properties of the
$\phi^{\text{inc}}_{i\alpha}$ solutions, the function $g^{\text{out}}_{i\alpha}(x')$ given in Eq.~\eqref{eq_gout2} becomes
\begin{equation}\label{eq:gout_def_2}
  g^{\text{out}}_{i\alpha}(x') = c_\alpha \phi^{\text{sct}}_{i\alpha}(x').
\end{equation}
Comparing Eqs.~\eqref{eq:gout_def_1} and \eqref{eq:gout_def_2}, we find

\begin{equation}\label{eq:Mii_matrix_elements}
  M^{ii}_{\alpha \beta} = - c_\alpha 
  \big( \mathbb{T}^i\big)^{-1}_{\alpha \beta}.
\end{equation}
We will use $M^{ii}$ and $M^{ij}$ given in Eqs.~\eqref{eq:Mii_matrix_elements} and \eqref{eq:Mii_matrix_elements}, respectively, to calculate the interface partition function $\mathcal{Z}_{\text{int}}$ given in Eq.~\ref{eq:partition-int-gij}. In order to calculate the {\it total} partition function, we need to calculate the colloid partition function, which we will present in the next section.

\subsection{Colloid Part}
In order to find the colloid partition function given in
Eq.~\eqref{eq:partition-col}, we have to calculate the boundary
Hamiltonian presented in Eq.~\eqref{eq:total-hamiltonian}.  Without
explicitly defining the boundary Hamiltonian, it is not possible to
evaluate and obtain a closed form expression. However, at equilibrium,
the boundary Hamiltonian is stable with respect to small fluctuations
for the field $f_i$, and thus we expand expand it in even powers of
the field $f_i$. Using the expansion of $f_i$ in terms of the incident
solutions given in Eq.~\eqref{eq:f_expansion}, to the lowest order we
can write the boundary Hamiltonian as
\begin{equation}\label{eq:Hb_matrix_elements}
  H^i_b[f_i] \approx \frac{\sigma}{2} \sum_{\alpha} 
  P_{i\alpha} H^i_{\alpha \beta} P_{i \beta}.
\end{equation}
Using Eq.~\eqref{eq:Hb_matrix_elements}, we are able to calculate the
colloid partition function Eq.~\eqref{eq:partition-col} as explained
in the next section.


\subsection{Total Partition Function}
After expanding the $\psi$ and $f$ fields based on Eqs.~\eqref{eq:f_expansion} and \eqref{eq:psi_expansion}, and using the $M$-matrix elements given in Eqs.~\eqref{eq:Mij_matrix_elements} and
\eqref{eq:Mii_matrix_elements} and the boundary Hamiltonian elements presented in Eq.~\eqref{eq:Hb_matrix_elements}, the total
partition function becomes a Gaussian integral over the expansion
coefficients $\Psi_{i\alpha}$ and $P_{i\alpha}$
\begin{multline}\label{eq:total-partition-function}
  \mathcal{Z} = \int \prod_{i \alpha} \dif \Psi_{i\alpha} \dif P_{i\alpha}
  \exp \bigg[
    - \frac{\sigma}{2k_B T} \\
    \begin{pmatrix} \Psi_{i\alpha} & P_{i\alpha} \end{pmatrix} 
    \mathbf{M}^{ij}_{\alpha \beta}
    \begin{pmatrix} \Psi_{j\beta} \\ P_{j \beta} \end{pmatrix}
    \bigg],
\end{multline}
with repeated index implying summation, and the $\mathbf{M}$ matrix
given by
\begin{equation}\label{eq:Mij_full_matrix_def}
  \mathbf{M}^{ij}_{\alpha \beta} = 
  \begin{pmatrix} c_\alpha \mathbb{U}^{ij}_{\alpha\beta} & 0 \\
    0 & 0 \end{pmatrix},
\end{equation}
for $i\ne j$ and
\begin{equation}\label{eq:Mii_full_matrix_def}
  \mathbf{M}^{ii}_{\alpha \beta} = 
  \begin{pmatrix} -c_\alpha \big(\mathbb{T}^i\big)^{-1}_{\alpha \beta} &
    \imath \delta_{\alpha \beta} \\
    \imath \delta_{\alpha \beta} & H^{i}_{\alpha \beta} \end{pmatrix},
\end{equation}
for $i=j$.
The partition function can then be calculated by performing the
Gaussian integral in Eq.~\eqref{eq:total-partition-function} and is
given in terms of the determinant of the $\mathbf{M}$ matrix
\begin{equation}
\label{eq:partition-final}
{\mathcal Z}={{\mathcal Z}''_0}^{-1} \,{\rm det}({\bf M})^{-\frac{1}{2}}\,,
\end{equation}
where ${\mathcal Z}''_0$ contains the normalization factor with
the constant parameters obtained from the Gaussian integral in
Eq.~\eqref{eq:total-partition-function}. In the next section, we will use the partition function, Eq.~\eqref{eq:partition-final} to calculate the Casimir energy.
 

\section{The Casimir Energy}
\label{sec:casimir-energy}
 
The Casimir energy at temperature $T$ is obtained by using the
Helmholtz relation,
\begin{equation}\label{eq:casimir-energy}
  \frac{\mathcal E}{k_B T}=
  -\ln\left(\frac{\mathcal{Z}}{{\mathcal Z}_\infty}\right)\,,
\end{equation}
with ${\mathcal Z}_\infty$ the partition function with all
colloids taken to infinite separation. 
Inserting Eq.~\eqref{eq:partition-final} in
Eq.~\eqref{eq:casimir-energy}, we find
\begin{equation}\label{eq:energy-semifinal}
 \frac{\mathcal E}{k_B T}= \frac{1}{2}\ln \text{det}
 \big(\mathbf{M}_\infty^{-1}\mathbf{M}\big) \,,
\end{equation}
with $({{\mathbf M}_{\infty}})^{ij}=\delta_{ij} {\mathbf M}^{ii}$
containing only the self-interacting terms. The elements of
the matrix $\mathbf{M}_\infty^{-1}\mathbf{M}$ can be explicitly written as
\begin{multline}
\big({\bf M}^{-1}_{\infty} {\bf M}\big)^{ij}_{\alpha \beta} =
 \mathbf{1}\delta_{ij}\delta_{\alpha \beta} + (1-\delta_{ij}) \\
 \times
 \begin{pmatrix}
   -c_\alpha \big(\mathbb{T}^i\big)^{-1}_{\alpha \gamma} &
   \imath \delta_{\alpha \gamma} \\ 
   \imath \delta_{\alpha \gamma} & H^i_{\alpha \gamma}
   \end{pmatrix}^{\!-1} \!\!\!\!
 \begin{pmatrix}
   c_\alpha \mathbb{U}^{ij}_{\gamma\beta} & 0 \\ 0 & 0 
 \end{pmatrix},
\end{multline}
with the explicit $2\times 2$ matrix indicating the $\Psi$ or $P$
components and the other matrix components indicated by the $i,j$ and
$\alpha,\beta$ indexes. The block-wise inverse of the $2\times 2$ matrix
can be taken and combined with the $2\times 2$ identity to give a
matrix that is lower block triangular
\begin{multline}
  \label{eq:Mmatrix_lower_tria}
\big({\bf M}^{-1}_{\infty} {\bf M}\big)^{ij}_{\alpha \beta} =\\
  \begin{pmatrix}
    \delta_{ij}\delta_{\alpha\beta} - 
    (1-\delta_{ij}) \widetilde{\mathbb{T}}^i_{\alpha \gamma} 
    \mathbb{U}^{ij}_{\gamma \beta} &
    0 \\
    X & \delta_{ij}\delta_{\alpha \beta}
  \end{pmatrix}.
\end{multline}
Note that the $X$ term is non-zero, but not important because the
determinant of a block triangular matrix is given by the product of
the determinants of the blocks along the diagonal.

Inserting Eq.~\eqref{eq:Mmatrix_lower_tria} into
Eq~\eqref{eq:energy-semifinal} gives
\begin{equation}
 \label{eq:energy-final}
 \frac{\mathcal E}{k_BT}=\frac{1}{2}\ln{\rm det}({\mathbf 1}-\widetilde{\mathbf T}{\mathbf U})\,,
\end{equation}
with ${\mathbf U}_{ij}={\mathbb U}^{ij}(1-\delta_{ij})$, and
$\widetilde{\mathbf T}_{ij}=\widetilde{\mathbb T}^{i}\delta_{ij}$.
The modified $T$-matrix, $\widetilde{\mathbb T}^i$,  comes from the upper
left corner of the block-wise inversion of the
$\mathbf{M}^{ii}_{\alpha\beta}$ matrix presented in
Eq.~\eqref{eq:Mii_full_matrix_def}, and is given by
\begin{equation}\label{eq:tmatrix-modified}
\widetilde{\mathbb T}^i= 
{\mathbb T}^i - {\mathbb T}^i \big[C\,{\mathbb T}^i\,C^{-1}+
C\,H^i\big]^{-1}{\mathbb T}^i\,,
\end{equation}
where $\mathbb{T}$ is the Dirichlet $T$-matrix matrix, $H^i$ is the
boundary Hamiltonian matrix elements defined in
Eq.~\eqref{eq:Hb_matrix_elements}, and $C$ is a diagonal matrix made
of the $c_\alpha$ coefficients from the Green's function expansion in
Eq.~\eqref{eq:greens_function_def_1}.

The general energy expression in Eq.~\eqref{eq:energy-final} gives the
fluctuation-induced interaction energy between $N$ colloidal particles
at a fluid interface.  We use the general energy expression to find
the interaction energy between two and three colloids.
\subsection{Interaction between Two Particles}
\label{sub:2particles}
For two colloids, the Casimir energy given by Eq.~\eqref{eq:energy-final}
reduces to
\begin{equation}
\label{eq:energy_two_particles} 
{\mathcal E}_{12} = \frac{k_B T}{2} \ln\det
\begin{pmatrix} \mathbf{1} & 
  \widetilde{\mathbb{T}}^1\mathbb{U}^{12} \\
  \widetilde{\mathbb{T}}^2\mathbb{U}^{21} &{\mathbf 1}
\end{pmatrix}.
\end{equation}
Using the formula for the determinant of a block matrix the two
particle energy takes on the familiar form
\begin{equation}
\label{eq:energy_two_particles} 
{\mathcal E}_{12} = \frac{k_B T}{2} \ln\det\big(\mathbf{1}-
  \widetilde{\mathbb{T}}^1\mathbb{U}^{12}
  \widetilde{\mathbb{T}}^2\mathbb{U}^{21}
\big).
\end{equation}
%
%
Note that Eq.~\eqref{eq:energy-final} is similar to the QED
Casimir energy \cite{universal} with the $T$-matrix ${\mathbb{T}}$ replaced with the
modified $T$-matrix $\widetilde{\mathbb{T}}$.

\subsection{Interaction between Three Particles}
\label{sub:3particles}
The full interaction energy given by Eq.~\eqref{eq:energy-final} for
three colloids can be written explicitly as
\begin{equation}
\label{eq:energy_three_particles} 
{\mathcal E} = \frac{k_B T}{2} \ln\det
\begin{pmatrix} \mathbf{1} & 
  \widetilde{\mathbb{T}}^1\mathbb{U}^{12} &
  \widetilde{\mathbb{T}}^1\mathbb{U}^{13} \\
  \widetilde{\mathbb{T}}^2\mathbb{U}^{21} &
  {\mathbf 1} &
  \widetilde{\mathbb{T}}^2\mathbb{U}^{23} \\
  \widetilde{\mathbb{T}}^3\mathbb{U}^{31} &
  \widetilde{\mathbb{T}}^3\mathbb{U}^{32} &
  {\mathbf 1}
\end{pmatrix}.
\end{equation}
Applying the properties of determinants for block matrices, this expression can be further simplified as
\begin{multline}
  \label{eq:energy_three_particle}
  \frac{\mathcal{E}}{k_B T} = 
  \frac{1}{2}\ln\det\big(\mathbf{1}-\mathbb{N}^{121}\big)+
  \frac{1}{2}\ln\det\big(\mathbf{1}-\mathbb{N}^{131} \big)\\
  +\frac{1}{2}\ln\det\big[{\mathbf 1} -
    ({\mathbf 1}-\mathbb{N}^{313})^{-1} 
    (\widetilde{\mathbb T}^3{\mathbb U}^{32}+\mathbb{N}^{312})\\
    \times({\mathbf 1}-\mathbb{N}^{212})^{-1} 
    (\widetilde{\mathbb T}^2{\mathbb U}^{23}+\mathbb{N}^{213})\big]\,,
\end{multline}
with the $N$-matrix defined as
\begin{equation}
  \mathbb{N}^{ijk} = 
  \widetilde{\mathbb{T}}^i\mathbb{U}^{ij}
  \widetilde{\mathbb{T}}^j\mathbb{U}^{jk}.
\end{equation}
The first two terms of Eq.~\eqref{eq:energy_three_particle} can be
identified as the two body energies between colloids 1-2 and 1-3
respectively. The third term particularly demonstrates that
fluctuation-induced interactions are not pairwise additive
\cite{emigScalar}.  
More specifically, Eq.~\eqref{eq:energy_three_particle} shows that the interaction
between colloids 2 and 3 has a more complicated form and contains the
indirect scattering of colloids 2 and 3 through the colloid 1.

The full three particle interaction energy can be considered as the
sum of the two body interactions plus a three-body interaction term
\begin{equation}
\label{eq123}
 \mathcal E = \sum_{i<j=1}^3{\mathcal E}_{ij} + {\mathcal E}_{123},
\end{equation}
with ${\mathcal E}_{123}$ the three body interaction, which can be evaluated by the
last term in Eq.~\eqref{eq:energy_three_particle} and subtracting the
two-body interaction between colloids 2 and 3
\begin{multline}
  \label{eq:energy_123}
  \frac{\mathcal{E}_{123}}{k_B T} = 
  \frac{1}{2}\ln\det \big[{\mathbf 1}
    -({\mathbf 1}-\mathbb{N}^{313}
    )^{-1} 
    (\widetilde{\mathbb T}^3{\mathbb U}^{32}+\mathbb{N}^{312}
    )\\\times({\mathbf 1}-\mathbb{N}^{212}
    )^{-1} 
    (\widetilde{\mathbb T}^2{\mathbb U}^{23}+\mathbb{N}^{213}
    )\big]-\\
  \frac{1}{2}\ln\det\big(\mathbf{1}-\mathbb{N}^{232}
  \big)
  \,.
\end{multline}
In the next section, we will employ Eqs.~\eqref{eq:energy_two_particles} and \eqref{eq:energy_three_particle} to calculate fluctuation-induced interactions between two and three spherical janus particles, respectively.


\section{Applications: circular Janus colloids}
\label{sec:applications}
We now apply the scattering formalism to calculate the fluctuation-induced interactions between
spherical Janus colloids. We obtain the
solutions to the helmholtz equation, Eq.~\eqref{eq:helmholtz}, for spherical particles in Subsec.~\ref{sec:2-sph-colloids}, which we use to find the matrices
${\mathbb T}$ and ${\mathbb U}$ for janus spherical partices.  In Subsec.~\ref{subsec:2particles}
we calculate the Casimir interaction between two spherical janus particles, and in Subsec.~\ref{sec:3-particles}
we obtain the fluctuation-induced interaction between three janus particles.
 
\subsection{Capillary wave solutions}
\label{sec:2-sph-colloids}
We consider spherical particles with the radius $R$, trapped at a
fluid interface with surface tension $\sigma$.  We assume that the
colloids are of Janus type and the contact line is pinned to them
with the contact angle $90^{\degree}$.  
The solutions to the Helmholtz equation, Eq.~\eqref{eq:helmholtz}, associated with the capillary
wave Hamiltonian in polar
coordinates read
\begin{eqnarray}
\label{eq:spherical-solution}
\phi^{\rm inc}_m &=& I_m(r/\lambda_{\rm c})e^{i m \theta}\,,\notag\\
\phi^{\rm sct}_m &=& K_m(r/\lambda_{\rm c}) e^{i m\theta}\,,
\end{eqnarray}
with $I_m(x)$ and $K_m(x)$ modified Bessel functions of the first
and second kinds, respectively. Note that $I_m(x)$ is regular as $x\to 0$ and $K_m(x)$
as $x\to\infty$. The free Green's function can then be written as
\begin{multline}
  G(r,\theta, r',\theta') = \\
  \sum_{m} I_m(r_</\lambda_c)e^{im\theta}
  K_m(r_>/\lambda_c)e^{-im\theta'},
\end{multline}
where $r_{<}$($r_>$) is the lesser (greater) of $r$ and $r'$. { This
form of the Green's function implies that all $c_m$'s are exactly
1.} In the following three subsections, we use these solutions to find
closed form expressions for the translation matrix $\mathbb{U}$, the
$T$-matrix $\mathbb{T}$, and the boundary Hamiltonian $H$.

\subsubsection{Translation Matrix}
To derive the translation matrix in polar coordinates, we will
expand the scattered solutions of the Helmholtz equation in terms of the incident
solutions.  In particular, we use Graf's addition theorem for Bessel functions
\cite{bessels_watson},
\begin{multline}
 \label{eq:graffs-addition}
 K_m(r_i/\lambda_{\rm c}) e^{\imath m\theta_i}= \\
 \sum_{m'=-\infty}^{\infty} \left[(-1)^{m'} e^{\imath(m-m')\theta_{ij}} 
   K_{m-m'}(d/\lambda_{\rm c})\right] \\
 \times I_{m'}(r_j/\lambda_{\rm c}) e^{\imath m' \theta_j}\,,
 \end{multline}
where $(d,\theta_{ij})$ are the polar coordinates of the vector that
connects the center of colloid $i$ to colloid $j$. Combining
Eqs.~\eqref{eq:translate_expansion} and \eqref{eq:spherical-solution} and comparing it with Eq.~\eqref{eq:graffs-addition}, we find the translation matrix in polar
coordinates
\begin{equation}
\label{eq:translation}
 {\mathbb U}^{ij}_{mm'} = (-1)^{m'} e^{\imath(m-m')\theta_{ij}}
 K_{m-m'}(d/\lambda_{\rm c})\,.
\end{equation}
We note that in most physical situations, $\lambda_{\rm c} = \sqrt{{\sigma}/{(\Delta \rho g})}$ is much
larger than any other length scale in the problem. In the limit
$d/\lambda_{\rm c}\ll 1$, the translation matrix given in Eq.~\eqref{eq:translation} for
$m=m'$ becomes
\begin{eqnarray}
\label{eq:translation-m=m'}
 {\mathbb U}^{ij}_{mm}=  {\mathbb U}^{ji}_{mm} 
 \approx (-1)^{m}\ln({2\lambda_{\rm c}}/{d})\,.
\end{eqnarray}
and for $m\ne m'$ is
\begin{multline}
\label{eq:translation-mnm'}
      {\mathbb U}^{ij}_{mm'} \approx 
      \frac{(-1)^{m'}}{2}\frac{|m-m'|!}{|m-m'|} \\ 
      \times e^{\imath(m-m')\theta_{ij}}
      \left(\frac{2\lambda_{\rm c}}{d}\right)^{|m-m'|}\,.
\end{multline}
We will use Eqs.~\eqref{eq:translation-m=m'} and \eqref{eq:translation-mnm'} along with the asymptotic forms of 
the $T$-matrix calculated in the next section, to obtain the asymptotic form of the Casimir energy, later in the paper. 
\subsubsection{Dirichlet Scattering Matrix}
We now plug the solutions of Helmholtz equation given in Eq.~\eqref{eq:spherical-solution} into
Eq.~\eqref{eq:scat_matrix_def} at the radius $R_i$ of the circular intersection of the object with the fluid interface and find
\begin{equation}\label{phim-sph}
 I_m(R_i/\lambda_{\rm c}) e^{\imath m \theta}+
 \sum_{m'}\mathbb{T}^i_{mm'} K_{m'}(R_i/\lambda_{\rm c})
 e^{\imath m'\theta}=0,
\end{equation}
which can be further simplified using the orthogonality of the exponentials over $\theta$ and it becomes
\begin{equation}
\label{eq:t-matrix_sphere}
 \mathbb{T}^i_{m}=-
 \frac{I_m(R_i/\lambda_{\rm c})}{K_m(R_i/\lambda_{\rm c})}\,.
\end{equation}
 Note that
the $T$-matrix, $\mathbb{T}$, is diagonal in $m$
($\mathbb{T}^i_{mm'}=\delta_{mm'}\mathbb{T}^i_m$).  In the limit
$R/\lambda_{\rm c}\ll 1$, the $T$-matrix for $m=0$ reads
\begin{equation}
 \label{eq:t-matrix-limit0}
 \mathbb{T}^i_0 \approx -\ln^{-1}(2\lambda_{\rm c}/R_i)\,,
\end{equation}
and for $m\ne 0$ is
\begin{equation}
 \label{eq:tmatrix-limit-n0}
 \mathbb{T}^i_m \approx -2\frac{|m|}{|m|!^2}
 \left(\frac{R_i}{2\lambda_{\rm c}}\right)^{2|m|}\,.
\end{equation}
We emphasize that the $T$-matrix of spherical particles is the same for any
particle with circular cross-section at the fluid interface,
such as disks and oblate spheroids.

\subsubsection{Boundary Hamiltonian} \label{sectionbh}
The boundary Hamiltonian in Eq.~\eqref{eq:boundary_hamiltonian}
contains three terms: a surface term from the integration by parts
from Eq.~\eqref{eq:colloid_hamiltonian-mid}, a term for the change in
the projected surface area, and a term for the change in the surface
area of the colloid in contact with each phase.

In the small gradient expansion the first two terms cancel
exactly. For a stiff Janus particle the area of the particle
surrounded by the contact line $f_i$ is fixed. If we extend the
contact line field into the interior of the colloid such that it
satisfies Laplace's equation, then the equal area expression can be
written
\begin{equation}
   \int_{\Omega_i} \!\!\! \dif^2 x\; \sqrt{1+(\nabla f_i)^2} =
   \int_{\Omega_i^{\text{ref}}} \!\!\!\! \dif^2 x\,.
\end{equation}
Performing a small gradient expansion on the first term and reordering
the terms we get the expression
\begin{equation}
  \Delta \Omega_i[f_i] = \frac{1}{2}\int_{\delta \Omega_i} \!\!\!\! \dif x\; 
  f_i \hat{n}\cdot\nabla f_i - \frac{1}{2}\int_{\Omega_i} \!\!\! \dif^2 x \;
  f_i \nabla^2 f_i.
\end{equation}
The first term given as an integral over the edge of the colloid will
exactly cancel the surface term from the boundary Hamiltonian in
Eq.~\eqref{eq:boundary_hamiltonian}. The second term is exactly zero
because we defined the $f_i$ field in the interior such that it
satisfies Laplace's equation.

Because of the cancellation of the first two terms in the boundary
Hamiltonian for stiff Janus particles, the boundary Hamiltonian is
given completely by the term with the interaction energies between the
colloid surface and the fluid phases. Janus particles are particles
whose surface has two distinct regions with different physical
properties, such as having a hydrophobic and hydrophilic
regions. These regions act to pin the contact line to a given position
on the particle. Mathematically this means that the
boundary Hamiltonian is zero if the contact line
matches the interface between the two regions of the Janus particle,
and it is very large (essentially infinite) otherwise.
\begin{table}
\begin{center}
\begin{tabular}{|c|c|}
  \hline
  Colloid Type & Boundary Ham. \\ \hline \hline
  Fixed & $H^i_{m} = \infty$ for all $m$. \\ \hline
  \multirow{2}{*}{Bobbing} & $H^i_0 = 0$ \\
  & $H^i_m = \infty$ for $m\ne 0$. \\ \hline  
  \multirow{3}{*}{Bobbing and Tilting} & $H^i_0 = 0$ \\
  & $H^i_{\pm 1} = 0$ \\
  & $H^i_{m}=\infty$ for $m\ne -1,0,1$ \\ \hline
\end{tabular}
\caption{The boundary Hamiltonian for a spherical Janus particle
  depends upon the type of fluctuations the colloid is allowed to
  undergo. This table lists the values of the boundary Hamiltonian
  depending on colloid fluctuation type.\label{tab:boundary_ham}}
\end{center}
\end{table}
The choice of whether to use zero or infinity depends on which types
of fluctuations the colloid is allowed to undergo. If the colloid is
fixed, then all fluctuations are forbidden. If the colloid is allowed
to bob up and down then the $m=0$ fluctuations are allowed, but all
other fluctuations are still forbidden. Finally if the colloid is
allowed to both bob up and down and tilt in any direction, then
the $m=0$ and $m=\pm1$ fluctuations are allowed. The values of the
boundary Hamiltonian for the various types of colloids are listed in
Tbl.~\ref{tab:boundary_ham}.

Using the two possible values for the boundary Hamiltonian, the
modified $T$-matrix for spherical particles can be found using
Eq.~\eqref{eq:tmatrix-modified}
\begin{equation}\label{eq:mod-tmat-sph}
  \widetilde{\mathbb{T}}^i_{m} = \begin{cases} 0 &
    \quad\text{if } H^i_m=0, \\
    \mathbb{T}^i_m & \quad\text{if } H^i_m=\infty,
  \end{cases}
\end{equation}
where $\mathbb{T}$ is the unmodified $T$-matrix given in
Eq.~\eqref{eq:t-matrix_sphere}.  The Casimir interaction for fluctuating
particles can be calculated using Eqs.~ \eqref{eq:translation},
\eqref{eq:t-matrix_sphere}, and \eqref{eq:mod-tmat-sph}.

\subsection{Two Spherical Particles}
\label{subsec:2particles}
As shown in Section \ref{sub:2particles}, the Casimir energy between two particles can be calculated using
Eq.~\eqref{eq:energy_two_particles}. Employing the Sylvester's
determinant theorem ${\rm det}({\bf 1-AB}) = {\rm det}({\bf 1-BA})$
and the fact that the $T$-matrix is diagonal, we can write the Casimir energy between two particles as %
\begin{equation}
 \label{eq:energy_2particles}
 \frac{{\mathcal E}}{k_B T}=\frac{1}{2}\ln{\rm det} 
 \big({\bf 1}-{\mathbb N}^{12}\big)\,,
\end{equation}
with the ${N}$-matrix given by 
\begin{equation}
 \label{eq:Nmatrix-conditioned}
 \mathbb{N}^{12}_{mm'} = \sum_{m''} 
 \mathbb{D}^{12}_{mm''}\mathbb{D}^{21}_{m''m'}\,.
\end{equation}
The ${D}$-matrix is the translation matrix sandwiched between the square root of
two modified ${T}$-matrices
\begin{equation}
  \label{eq:Dmatrix}
  \mathbb{D}^{ij}_{mm'} = \sqrt{\widetilde{\mathbb{T}}^i_m}
  \mathbb{U}^{ij}_{mm'}\sqrt{\widetilde{\mathbb{T}}^j_{m'}}\,.
\end{equation}
The form of ${N}$-matrix given in
Eq.~\eqref{eq:Nmatrix-conditioned} is more convenient for both asymptotic analysis and
numerical calculations. Considering that the capillary length is the largest length scale
in the system and plugging Eqs.~\eqref{eq:tmatrix-limit-n0} and
\eqref{eq:translation-mnm'} into Eq.~\eqref{eq:Dmatrix}, we find the ${D}$-matrix
in the limit of $\lambda_c \gg R,d$
\begin{equation}
  \label{eq:Dmatrix_asym}
\mathbb{D}^{ij}_{mm'} \propto 
\bigg(\frac{R_i}{2\lambda_{\rm c}}\bigg)^{|m|}
\bigg(\frac{R_j}{2\lambda_{\rm c}}\bigg)^{|m'|}
\bigg(\frac{2\lambda_{\rm c}}{d}\bigg)^{|m-m'|}\!\!\!.
\end{equation}
Careful consideration of the exponents shows, that in the limit of
large capillary length ($\lambda_{\text{c}} \to \infty$) only the terms in which
$m$ and $m'$ have different signs remain non-zero.  If we plug
Eq.~\eqref{eq:Dmatrix_asym} into Eq.~\eqref{eq:Nmatrix-conditioned}
and assume that $R_i=R_j=R$, the ${N}$-matrix elements can be
written as a power series in $R/d$
\begin{equation}
  \label{eq:Nmatrix_power_series}
  \mathbb{N}_{mm'}=\sum\limits_{m''=0}^{\infty} A^{m m'}_{m''} 
  \bigg(\frac{R}{d}\bigg)^{2m''+|m+m'|},
\end{equation}
where the $A^{mm'}_{m''}$ coefficients can contain logarithmic functions
in terms of $R$, $d$, and $\lambda_{\rm c}$.



In the following we use the ${N}$-matrix given in
Eq.~\eqref{eq:Nmatrix_power_series} to obtain the large separation
asymptotic energies between the spherical Janus colloids for three
different cases: colloids frozen at a vertical position; bobbing
colloids that fluctuate only vertically with the interface; and
bobbing and tilting colloids that both fluctuate vertically and tilt
side to side.

\subsubsection{Fixed Janus colloids}
To calculate the Casimir interaction between two frozen janus
particles at large separations, we need to find the asymptotic form of
$\ln{\rm det}({\bf 1}-\mathbb{N})$.  Since the determinant is
invariant upon even interchanging of the rows and columns, the $N$-matrix 
can be written as
\begin{equation}
\label{eq:Nmatrix-interchanged}
\mathbb{N} = \left(
\begin{array}{c|cccccc}
 \mathbb{N}_{00} & \mathbb{N}_{01} & \mathbb{N}_{0-1} & \dots \\ \hline
 \mathbb{N}_{10} & \mathbb{N}_{11} & \mathbb{N}_{1-1} & \dots \\ 
 \mathbb{N}_{-10} & \mathbb{N}_{-11} & \mathbb{N}_{-1-1} & \dots \\
 \vdots & \vdots & \vdots & \ddots \\ 
  \end{array}
\right)\,.
\end{equation}
The lines in Eq.~\eqref{eq:Nmatrix-interchanged} divide the $N$-matrix
 into four blocks. Simplifying the determinant for block
matrices gives
\begin{multline}\label{eq:logdet-fixed}
  \ln{\rm det}({\bf 1}-\mathbb{N}) = 
  \ln(1-\mathbb{N}_{00})+\ln{\rm det}({\bf 1}-{\bf A}) \\
  +\ln{\rm det}({\bf 1}-\frac{({{\bf 1}-{\bf A}})^{-1} 
    {\bf B}}{1-\mathbb{N}_{00}} )\,,
\end{multline}
where $\mathbf{A}$ is the lower right block
%
%
and $\mathbf{B}$ is the matrix product of the lower left column and
upper right row in Eq.~\eqref{eq:Nmatrix-interchanged}.
%
%
%
Due to the scaling behavior of ${\mathbb N}_{mm'}$ given in
Eq.~\eqref{eq:Nmatrix_power_series}, the asymptotic behavior of the
Casimir energy for fixed particles is dominated by the first term in
Eq.~\eqref{eq:logdet-fixed}.  Furthermore, as discussed above, for
fixed colloids trapped at the fluid interface $\widetilde{\mathbb T}^i={\mathbb
  T}^i$.  Using the $T$-matrix and translation matrix ${\mathbb U}$ given in Eqs.~\eqref{eq:t-matrix-limit0} and
\eqref{eq:translation-m=m'} respectively, in the
limit $\lambda_{\rm c}/R\gg1$ the asymptotic energy reads
\begin{equation}
\label{eq:two-frozen-colloids}
\frac{\mathcal E}{k_B T}\approx \frac{1}{2}\ln\left(1-
\frac{\ln^2(2\lambda_{\rm c}/d)}
{\ln(2\lambda_{\rm c}/R_1)\ln(2\lambda_{\rm c}/R_2)}\right)\,.
\end{equation}
Note that Eq.~\eqref{eq:two-frozen-colloids} is in complete agreement with
the results presented Refs.~\cite{oettel06,oettel07}.

\subsubsection{Bobbing Janus colloids} \label{secbobonly}

For bobbing colloids we have $\widetilde{\mathbb T}_{0}=0$  as discussed in Section \ref{sectionbh}. This implies that $\mathbb{N}_{mm'}=0$ if either $m$ or $m'$ is equal
zero, see Eqs.~\eqref{eq:Nmatrix-conditioned} and \eqref{eq:Dmatrix}. Furthermore, using the identity $\ln \det(\mathbf{1}-\mathbb{N}) = -\sum_{s=1}^\infty \frac{1}{s}
  \operatorname{\mathrm{tr}}(\mathbb{N}^s)$ and the power series
relation for the ${N}$-matrix given in Eq.~\eqref{eq:Nmatrix_power_series}
we find that only the $s=1$ term
and the $m=\pm 1$ in Eq.~\eqref {eq:Nmatrix_power_series} contribute to the asymptotic Casimir energy in the leading order terms in $R/d$.  The Casimir energy can then be written as
\begin{equation}
  \label{eq:asymb}
  \frac{\mathcal E}{k_B T} \approx -\frac{1}{2}
  \big(\mathbb{N}^{12}_{11}+\mathbb{N}^{12}_{-1-1}\big). 
\end{equation}
Note that as explained in the paragraph above Eq.~\eqref{eq:Nmatrix_power_series}, only the terms in which the signs of $m$ and $m'$ are the same are different from zero.  Keeping only the $m''=1$ term in
Eq.~\eqref{eq:Nmatrix_power_series} results to
%
\begin{equation}
 \label{eq:energy-bob}
 \frac{\mathcal E}{k_B T}\approx - \frac{R^2_1R^2_2}{d^4}\,.
\end{equation}
Equation~\eqref{eq:energy-bob} is in agreement with the results given in
Refs.~\cite{oettel06,oettel07,eft_deserno1}. The power law in Eq.~\eqref{eq:energy-bob} will be modified if the particles can bob and tilt as presented in the next section.

\subsubsection{Bobbing and tilting Janus colloids}  \label{secbobtilt}

As was discussed in Section \ref{sectionbh}, for spherical Janus colloids that can both bob
and tilt the modified $\widetilde{\mathbb{T}}_m=0$ for $m=0$ and $m=\pm 1$.
Therefore, the $m=\pm2$ terms will be the leading
order terms in $R/d$ expansion of the Casimir energy and is given by
%
\begin{equation}
  \label{eq:asym-bt}
  \frac{\mathcal E}{k_B T} \approx 
  -\frac{1}{2}(\mathbb{N}^{12}_{22}
  +\mathbb{N}^{12}_{-2-2})\,.
\end{equation}
Note that the leading
contribution in the power series expansion of the $N$-matrix given in
Eq.~\eqref{eq:Nmatrix_power_series} comes from the $m''=2$ term. To this end, the
asymptotic energy at large separation can be written as
\begin{equation}
\label{eq:energy-two-colloids-tiltandbob}
\frac{\mathcal E}{k_B T} \approx -\frac{9\,R^4_1R^4_2}{d^8}\,.
\end{equation} 
Equation~\eqref{eq:energy-two-colloids-tiltandbob} is in agreement
with the results obtained in Refs.~\cite{oettel06,oettel07} for two spherical particles. 
In the next section, we will investigate the interaction between three spherical particles.

\subsection{Three Spherical Particles}
\label{sec:3-particles}
Similar to the case of two particle system, it is convenient to write the Casimir energy in terms of $D$-matrices 
given in Eq.~\eqref{eq:Dmatrix} such that each translation matrix $\mathbb U$ is sandwiched between the square root of two
$T$-matrices. Then, the three particle energy given
in Eq.~\eqref{eq:energy_123} can be written completely in terms of
$D$-matrices as
\begin{multline} \label{eq123for3}
  \frac{\mathcal{E}_{123}}{k_B T} =
  \frac{1}{2}\ln \det\Big(
  \big(1-\mathbb{D}^{32}\mathbb{D}^{23}\big)^{-1}
  \big(1-\mathbb{M}\big)\Big),
\end{multline}
with the ${M}$-matrix
\begin{multline}  \label{eqMfor3}
  \mathbb{M} = \big(1-\mathbb{D}^{31}\mathbb{D}^{13}\big)^{-1}
  \big(\mathbb{D}^{32}+\mathbb{D}^{31}\mathbb{D}^{12}\big) \\ \times
  \big(1-\mathbb{D}^{21}\mathbb{D}^{12}\big)^{-1}
  \big(\mathbb{D}^{21}+\mathbb{D}^{21}\mathbb{D}^{13}\big).
\end{multline}
Using the expression for the ${D}$-matrix given in
Eq.~\eqref{eq:Dmatrix_asym}, we will calculate the large separation
asymptotic interactions for the three particle system in the next section. Once again, we will consider three cases of frozen, bobbing only and bobbing and tilting colloids trapped at the fluid interface.



\subsubsection{Frozen particles}
\label{susub:3fixed}

For fixed colloids at the fluid interface, the ${M}$-matrix can be reordered similar
to the ${N}$-matrix in Eq.~\eqref{eq:Nmatrix-interchanged}. The leading
order contribution to the three body interaction, defined in Eqs.~\eqref{eq123} and  \eqref{eq:energy_123}, comes from the $m=m'=0$ component and is
\begin{equation}
  \frac{\mathcal{E}_{123}}{k_B T} \approx \frac{1}{2} \ln \bigg(
  \frac{1-\mathbb{M}_{00}}{
    1-\mathbb{D}^{23}_{00}\mathbb{D}^{32}_{00}} \bigg).
\end{equation}
Using the asymptotic expressions for the matrices $\mathbb{U}$ and  $\mathbb{T}$ given in
Eqs.~\eqref{eq:translation-m=m'} and \eqref{eq:t-matrix-limit0} respectively, the three body interaction can be written as 
\begin{equation}
\label{eq:three-fixed-en}
 \frac{{\mathcal E}_{123}}{k_B T} \approx \frac{1}{2} \ln\left[\frac{1-(g_{12}^2+g_{13}^2+g_{23}^2+2g_{12}g_{13}g_{23})}{(1-g_{12}^2)(1-g_{13}^2)(1-g_{23}^2)}\right]\,,
\end{equation}
with
\begin{equation}
 g_{ij} = \frac{\ln(2\lambda_{\rm c}/d_{ij})}{\sqrt{\ln(2\lambda_{\rm c}/R_{i})\ln(2\lambda_{\rm c}/R_{j})}}\,.
\end{equation}

\subsubsection{Bobbing only}
\label{susub:3bob}

As noted previously, for bobbing only colloids the modified
$\widetilde{\mathbb T}_0 = 0$. Once again, the logarithmic determinant in Eq.~\eqref{eq123for3} can be simplified using  the identity $\ln \det(\mathbf{1}-\mathbb{N}) = -\sum_{s=1}^\infty \frac{1}{s}
  \operatorname{\mathrm{tr}}(\mathbb{N}^s)$. The argument of the determinant itself can be expanded in powers of the $D$-matrix, see Eqs.~\eqref{eq123for3} and  \eqref{eqMfor3}, and it can be written up to $4^{\text{th}}$ order as
\begin{multline}
  1-\big(1-\mathbb{D}^{32}\mathbb{D}^{23}\big)^{-1}\big(1-\mathbb{M}\big) = \\
  \mathbb{D}^{32}\mathbb{D}^{21}\mathbb{D}^{13} +
  \mathbb{D}^{31}\mathbb{D}^{12}\mathbb{D}^{23} \\ +
  \mathbb{D}^{31}\mathbb{D}^{13}\mathbb{D}^{32}\mathbb{D}^{23} +
  \mathbb{D}^{32}\mathbb{D}^{21}\mathbb{D}^{12}\mathbb{D}^{23} \\ +
  \mathbb{D}^{31}\mathbb{D}^{12}\mathbb{D}^{21}\mathbb{D}^{13} + \cdots.
\end{multline}
Note that a matrix product of an odd number of $\mathbb D$-matrices will
always evaluate to zero in the limit of large capillary
length. The three remaining $4^{\text{th}}$ order terms can be
rewritten in terms of the $N$-matrices defined in
Eq.~\eqref{eq:Nmatrix-conditioned}
\begin{equation} \label{sec:3particlee123}
  \frac{\mathcal{E}_{123}}{k_B T} \approx
  -\frac{1}{2}\operatorname{tr}\big(
  \mathbb{N}^{13}\mathbb{N}^{32} +\mathbb{N}^{21}\mathbb{N}^{32} +
  \mathbb{N}^{12}\mathbb{N}^{13}\big).
\end{equation}
We showed in Sect.~\ref{secbobonly} that for the two particle system,
the leading order term in the interacting energy at large separation
is obtained from the $|m|=1$ terms given in
Eq.~\eqref{eq:Nmatrix_power_series}.  The three particle asymptotics
for bobbing Janus particles can now simply be read off from the two
particle asymptotics in Eq.~\eqref{eq:energy-bob}
\begin{equation}
  \label{eq:3-bobing-asym}
  \frac{\mathcal{E}_{123}}{k_B T} \approx -\bigg(
  \frac{R_1^4 R_2^2 R_3^2}{d_{12}^4d_{13}^4}+
  \frac{R_1^2 R_2^4 R_3^2}{d_{12}^4 d_{23}^4}+
  \frac{R_1^2 R_2^2 R_3^4}{d_{13}^4d_{23}^4}\bigg)\,.
\end{equation}
\begin{center}
\begin{figure}
\includegraphics[scale=1]{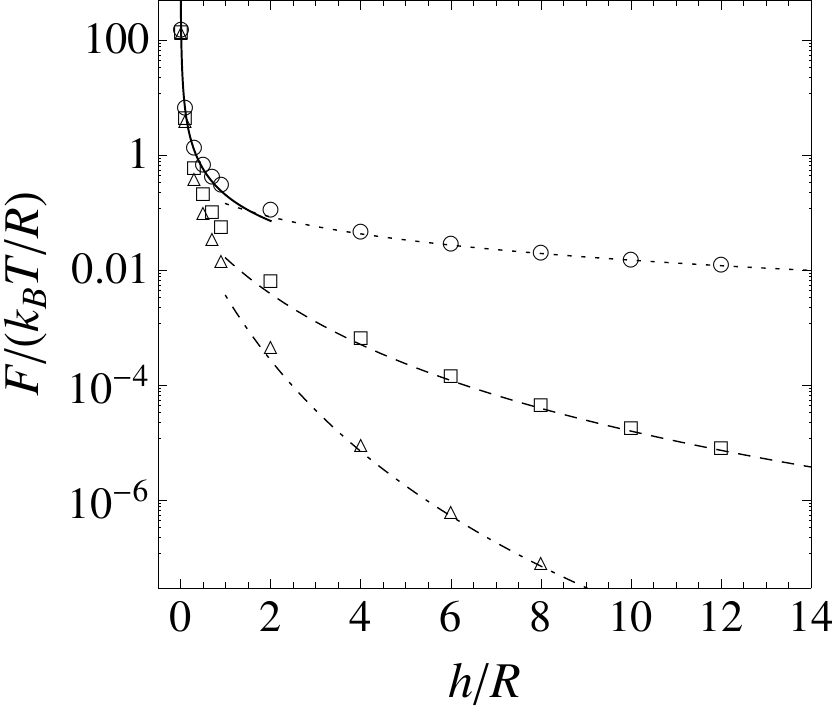} 
\caption{\label{fig:two-parts}Comparison of the
  numerical and asymptotic results. The plots show the
  fluctuation-induced interaction between two identical spherical Janus
  particles of radius $R$ as a function of surface-to-surface
  separation $h/R$. The circles, squares, and triangles illustrate the
  numerical results for the frozen, bobbing only, and bobbing and
  tilting colloids, respectively. The dotted, dashed, and dotted-dashed
  lines show the asymptotic interactions for the frozen, bobbing only, and bobbing and
  tilting colloids, respectively.  The solid curve at short
  separations shows the Casimir interaction obtained by Derjaguin
  approximation \eqref{eq:derj}.  }
\end{figure}
\end{center}

%

\subsubsection{Bobbing and tilting particles}

As explained in Subsec.~\ref{sub:2particles}, for the colloids that can undergo both bobbing and tilting fluctuations,
$\widetilde{\mathbb T}_0=\widetilde{\mathbb T}_{\pm 1}=0$; and thus for
calculating the asymptotics, the $|m|=2$ terms are the dominant ones in the sum given in 
Eq.~\eqref{eq:Nmatrix_power_series} for calculating $N$-matrix.
Analogous to bobbing colloids, the asymptotic energy for three bobbing and tilting Janus particles can be calculated using Eq.~\eqref{sec:3particlee123}
and is equal to
\begin{equation}
  \label{eq:three-tilt-en}
  \frac{\mathcal{E}_{123}}{k_B T} \approx -81 \!\left(\!
  \frac{R_1^8 R_2^4 R_3^4}{d_{12}^8d_{13}^8} \!+\!
  \frac{R_1^4 R_2^8 R_3^4}{d_{12}^8 d_{23}^8} \!+\!
  \frac{R_1^4 R_2^4 R_3^8}{d_{13}^8d_{23}^8}\right)\!.
\end{equation}
%

\section{Numerical Results}
\label{sec:numerics}

To calculate the Casimir energy numerically, we use the general energy
expression given in Eq.~\eqref{eq:energy-final}. The $m$ index for the translation 
$T$-matrices given in Eqs.~\eqref{eq:t-matrix_sphere} and \eqref{eq:translation} respectively,  is truncated at some $m_{\text{max}}$. The Casimir energy 
then becomes the logarithm of a large determinant that depends on the cutoff $m_{\text{max}}$. It can be shown numerically that
Casimir energy converges exponentially in
$m_{\text{max}}$. In our numerical calculations, convergence is assumed when
$m_{\text{max}}$ is increased by 5 but the
relative change in the Casimir energy is less than $10^{-4}$. 

Figure~\ref{fig:two-parts} shows the fluctuation-induced interactions
calculated numerically between frozen, bobbing only, and bobbing and
tilting spherical Janus particles. The dotted, dashed, and dotted-dashed lines correspond to the asymptotic
Casimir energies derived in
Eqs.~\eqref{eq:two-frozen-colloids}, \eqref{eq:energy-bob} and
\eqref{eq:energy-two-colloids-tiltandbob}, respectively. As shown in the figure,
there is a very good agreement at larger separations ($h/R >2$)
between the exact numerics and asymptotic results.  

Quite
interestingly, at short separations ($h/R < 0.1$), we find an
excellent agreement between our results and the Derjaguin or proximity
approximation \cite{derjaguin_approximation}.  The Derjaguin
approximation replaces the energy at close separations with the sum
over the Casimir interaction per length between parallel lines (along
the contact lines of the colloids).  The energy between two parallel
lines is calculated in Ref.~\cite{kardarlif2d} and the Derjaguin
approximation for the interaction between the two spherical colloids
at an interface is given by \cite{oettel06,oettel07}
\begin{equation}
\label{eq:derj}
F \approx -k_B T \frac{\pi^2}{48} \frac{R^{1/2}}{h^{3/2}}\,,
\end{equation}
with $h=d-2R$.

Note that the scattering method is extremely fast for the calculation of
fluctuation-induced interactions between two particles. A single data
point is produced in less than $0.01$ seconds at intermediate
separations $h/R \sim 1$ between two particles. For shorter
separations one needs to truncate all the matrices involved at larger
$m_{\text{max}}$ resulting in longer computation time. However, the
longest computation time needed for a single data point was less than
two minutes, for a separation of $h/R = 0.001$ with
$m_{\text{max}} \approx 200$. This is a signifigant speedup compared
to previous methods. More than 5000 points are needed to calculate the
interaction energy at the separation around $0.01R$ using the path
integral formalism, see for example Refs.~\cite{oettel06,oettel07}.

\begin{center}
\begin{figure}
\includegraphics[scale=0.95]{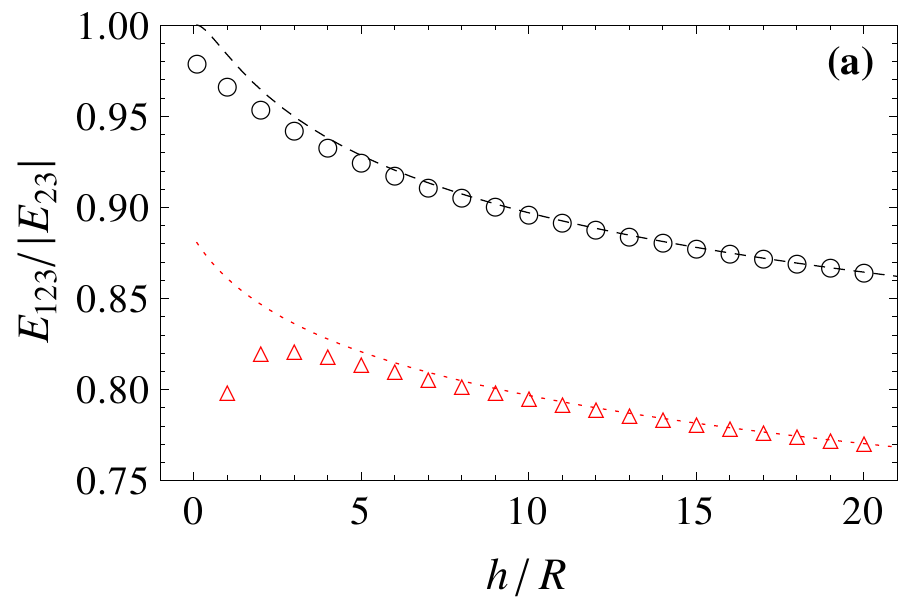} 
\includegraphics[scale=0.95]{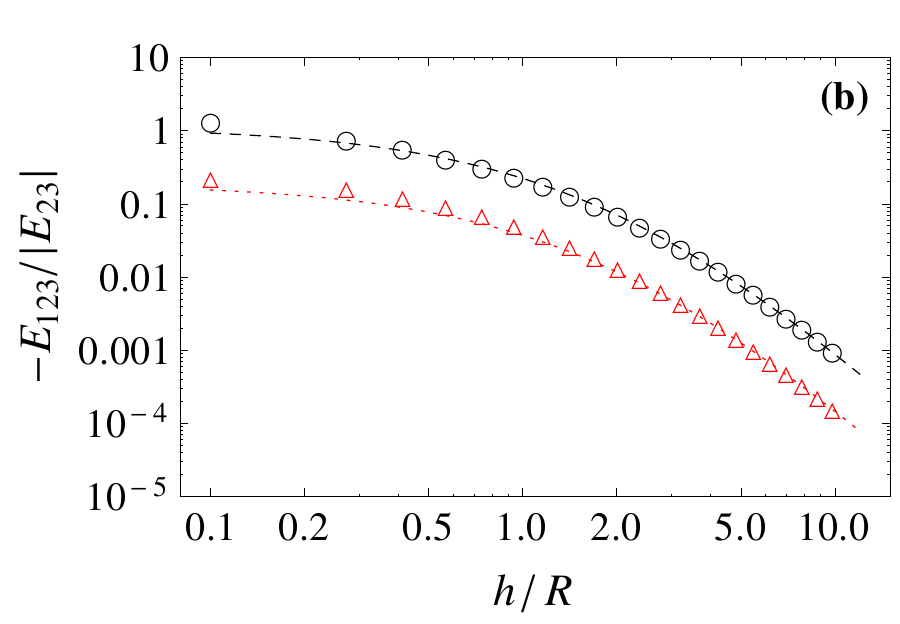} 
\includegraphics[scale=0.95]{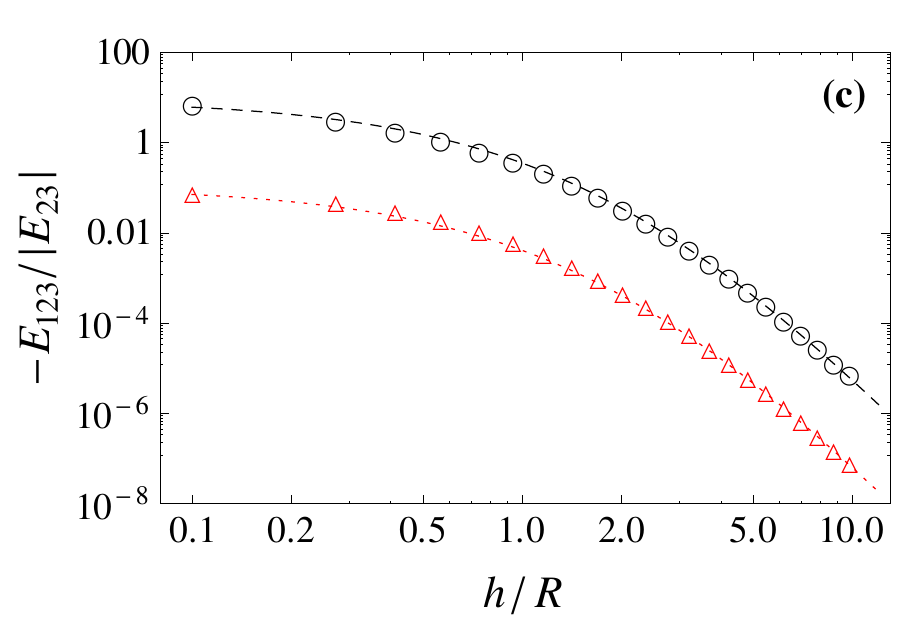} 
\caption{\label{fig:three-parts}(Color online) Comparison of the
  numerical and asymptotic results for three spherical Janus colloids.
  The plots show the ratio of the three body effect $E_{123}$ to the
  two body Casimir energy $E_{23}$ versus the distance between two
  colloids.  The empty circles and black lines represent the numerical and
  asymptotic energies, respectively, for the three particles sitting
  on a line ($d_{12}=d_{23}=d_{13}/2$). The empty triangles and
  dotted curve show the numerical and asymptotic energies, respectively,
  for three colloids sitting at the vertices of an equilateral
  triangle ($d_{12}=d_{23}=d_{13}/2$). 
}
\end{figure}
\end{center}

Figure~\ref{fig:three-parts}(a), (b) and (c) illustrate the three body
effect $\mathcal{E}_{123}$ normalized by the two body interaction
$\mathcal{E}_{23}$ between frozen, bobbing only and tilting and
bobbing colloids, respectively.  There is a very good agreement
between the asymptotics and exact numeric results for large separation,
as shown in the figure. The dashed lines and empty circles show the asymptotic results and the numerics for the three particles
sitting on a line at equal distance ($d_{12}=d_{23}=d_{13}/2$). The
dotted line and empty triangles illustrate three colloids sitting at
the vertices of an equilateral triangle
($d_{12}=d_{23}=d_{13}/2$). The asymptotic results for the three body
effect for frozen, bobbing only and bobbing and tilting spherical
colloids are obtained based on Eqs.~\eqref{eq:three-fixed-en},
\eqref{eq:3-bobing-asym} and \eqref{eq:three-tilt-en}, respectively.
The asymptotic energies for two colloids are presented in
Eq.~\eqref{eq:two-frozen-colloids} for frozen ,
Eq.~\eqref{eq:energy-bob} for bobbing only, and
Eq.~\eqref{eq:energy-two-colloids-tiltandbob} for bobbing and tilting
particles.
 
Quite interestingly we find that for the three body Casimir
interaction, the three body effect for fixed colloids is repulsive and
comparable to the two body interaction
energy. Figure~\ref{fig:three-parts}(a) shows that this is indeed true
for all separations.  On the other hand,
Figs.~\ref{fig:three-parts}(b) and 2(c) show that the three body
effect for bobbing only and bobbing and tilting colloidal particles is attractive such
that the total fluctuation-induced energy for three particles is
larger than the sum of three pairwise interactions.  Note that this
strengthening effect is negligible at large separations but quite
significant at short separations.

\section{Summary and Conclusions}
\label{sec:summary}

In this work, we extended the scattering formalism previously employed for the calculation of the Casimir forces in QED  \cite{roya10, ours1,rahi08,emig06,ours,Emig:2009fk,Graham:2010uq,Maghrebi:2011kx}, to obtain
the Casimir interaction between colloidal particles trapped at a fluid
interface. Since in soft matter systems, the colloids may also
fluctuate, we implemented the effect of boundary fluctuations in the
scattering method.  We found that the scattering matrix of a
fluctuating colloid can be written as a mixture of the scattering
matrix of frozen colloids and a boundary matrix determined by
the colloids fluctuations energy costs.  

The augmented scattering method has several advantages: (i) 
It can be easily applied to colloids with different geometries compared to other techniques, 
as in the scattering formalism one does not need to calculate the specific Green's function for each system separately. (ii) Inclusion of colloid fluctuations to the scattering method is developed quite generally for all geometries.
(iii) The numerical calculation of the fluctuation-induced forces is much faster 
using the present method. (iv) Finally, the method developed in this work is also applicable to many particle systems. 

To show the effectiveness of
the developed method, we applied it to a system of Janus spherical
colloids trapped at the interface between two fluid phases. For two
particles, we reproduced the well-known asymptotic results \cite{oettel06,oettel07}. 
For three particles, we found the asymptotic results and numerically obtained the three body effect.  Besides the excellent agreement between the
numerics and asymptotics, we found that the three body effect for
frozen particles at the interface is repulsive and decreases the total
free energy between three particles.  For fluctuating colloids, we
showed that the three body effect is attractive such that it
strengthens the total fluctuation-induced interaction. This increase
is much larger at short separations.
The authors would like to thank Mehran Kardar for useful
discussions. This work was supported by the National Science
Foundation through Grant No. DMR-06-45668 (R.Z.).

\bibliography{ref.bib}

\end{document}